\documentstyle[prc,aps,eqsecnum]{revtex}

\title{On an Extended PCAC Relation}
\author{S.Ying}
\address{CCAST (World Laboratory), P. O. Box 8730 \\Beijing 100080, China\\
and\\
Theory Group, Physics Department, Fudan University\\ Shanghai 200433,
China
\thanks{The present address.}}
\date{\today}

\tighten
\draft

\begin{document}

\maketitle

\begin{abstract}
{
  We explore a consistent way to extend the partially conserved axial
vector current (PCAC) relation and corresponding current algebra results
in two strongly correlated directions: 1) towards a search for a set of
systematic rules for the establishment of PCAC related relations in a
finite low momentum transfer region, and,  for the extrapolation of the
momentum
transfer $q^2$ to zero when deriving the low energy PCAC results that can be
compared to experimental data and 2) towards taking into account, besides the
conventional one, the only other possibility of the spontaneous chiral symmetry
breaking, $SU(2)_L\times SU(2)_R \to SU(2)_V$, inside a baryonic system by
a condensation (in the sense to be specified in the paper) of diquarks. The
paper includes investigations of a chiral Ward-Takahashi identity, the
explicit chiral symmetry breaking by a finite current quark mass, the
modification of the PCAC relation and its consequences. Two explicit
relations between the nucleon axial vector form factor $g_A(q^2)$,
pseudo-scalar form factor $g_P(q^2)$ and the pion-nucleon coupling
constant $g_{\pi NN}(q^2)$ is obtained. One of the relation is
confirmed, within the experimental error, by observation in the $0<-q^2 <
0.2$ $\mbox{GeV}^2$ region. The other one, which relates $g_A(q^2)$ and
$f_\pi(q^2) g_{\pi NN}(q^2)$ is studied by using known empirical
facts and dispersion relation. Theoretical uncertainties are discussed.
Certain inconsistencies, which
is in favor of the introduction of diquark condensation, is discovered.
We briefly discuss how diquark condensation could provide an answer to
the question of where about of the quark numbers in a nucleon and
a nucleus, which is raised in explaining puzzles observed in the
violation of Gottfried sum rule and EMC effects.
}
\end{abstract}


\section{Introduction}
\label{sec:Intro}

    Model independent partially conserved axial vector current (PCAC)
relation and
corresponding current algebra results are in conformity with experimental data
within a few percent. It is quite impressive compared to other
observables in strong interaction. This good agreement between theory
and experiments is interpreted as due to an underlying approximate
chiral $SU(2)_L\times SU(2)_R$
symmetry, which is explicitly broken down by up and down current quark masses
of a few MeV, much smaller than the hadronic mass scale of 1 GeV. The lightest
hadronic particle pions are considered as, in the limit that the current masses
of the up and down quarks vanish, the Goldstone bosons of a spontaneous
breaking down of the above mentioned symmetry induced by a non-vanishing
vacuum expectation value of the quark field bilinear operator $\bar\psi\psi$,
namely, $\mathopen{\langle 0\,|}\bar\psi\psi\mathclose{|\, 0\rangle}
\ne 0$. Equipped with new relevant empirical information,
we provide a refined analysis of the old results, which are based on an
extrapolation of the physical quantities
to $q^2\to 0$ \cite{AdlerBook}, to include a wider range of momentum transfer,
and, may be more interestingly, to explore possible extensions that are
observable and are nevertheless consistent with our present knowledge.

    The possibility of the formation of a superconducting phase in a massless
fermionic system, which is a realization of one of the possible phases in which
the chiral symmetry is spontaneously broken down to a flavor
(isospin) symmetry,
is investigated in Ref. \cite{YING1} base on a 4-fermion interaction
model. It was argued that the relativistic superconducting phase might be
relevant to the creation of baryons in the early universe due to its quantum
mechanical nature. If this scenario reflects the nature at least at a
qualitative level, it would be of interest to study the possible existence of
such a phase in the hadronic system at the present day condition.
Since it is unlikely that the present day strong interaction
vacuum at large scale is in the superconducting phase for reasons given in the
following, the only regions where the superconducting phase can be found are
inside a nucleon, a nucleus, the center region of a heavy ion collision,
and an astronomical object like a neutron star, a
quasar, etc. Albeit the possible superconducting phase in the baryon creation
era of the early universe may have had disappeared entirely at certain
previous time during the evolution of the universe, it is still a worthwhile
effort to search for such a phase at the present time. One
of the reasons is that if such a phase can be found, its properties can be
studied in domestic laboratories. Its existence is a reasonable possibility
since it is shown \cite{YING1} that
given suitable coupling constants, the phase in
which $\mathopen{\langle 0\,|}\bar\psi\psi\mathclose{|\, 0\rangle} \ne 0$
changes into a superconducting phase as the baryon (or quark) density is
raised. The empirical need for such an assumption will be discussed in
the conclusion parts of the paper rather than in this section because most of
the individual phenomenon considered has its own explanation in terms of
conventional picture with various degrees of success at the present
theoretical and/or experimental precisions. We are interested in a
search for a consistent explanation of a large set of observations. Motivated
by
these considerations, we explore
some of the phenomenological consequences of a possibility in which the
interior of a nucleon contains diquark condensation \cite{diquark}, which is
enhanced in a nucleus, that spontaneously breaks the chiral symmetry.

In order to differentiate the two hidden chiral (asymmetric)
phases, the phase in which $\mathopen{\langle \,}\bar \psi \psi
\mathclose{\,\rangle} \ne 0$ and no diquark condensation is called the
Nambu phase \cite{NB1} and the phase in which there is a
diquark condensation is called the superconducting phase in this paper.
The model dependency of the discussion is reduced as much as possible. For
that purpose, we classify the spontaneous chiral symmetry belonging to the
same chain, namely, $SU(2)_L\times SU(2)_R\to SU(2)_V$, into categories
characterized by their order parameters introduced in Ref.
\cite{YING1}, which are generic in nature. The Nambu
phase is characterized by a non-vanishing $\sigma$ and vanishing
$\phi^{c\mu}$ and $\bar \phi^\mu_c$. The superconducting phase is
characterized by non-vanishing $\phi^{c\mu}$ and $\bar\phi^{\mu}_c$ and
possibly a non-vanishing $\sigma$. The generality
of the discussion allows the results to  be applied to any model with
a similar phase structure as the one obtained in Ref. \cite{YING1}.

    The paper is organized in the following way. In section \ref{sec:WT}, a
chiral  Ward-Takahashi identity is studied by presenting the main
results, which establish the existence of the Goldstone diquark
excitation in the superconducting phase.
 Section \ref{sec:m0pert} deals with the perturbation of a small
current quark mass term in the Lagrangian. The mixing of the
Goldstone diquark excitation in the divergence of the hadronic
axial vector current operator in a baryonic system containing diquark
condensation is demonstrated. Under the assumption that there is
a diquark condensation inside a nucleon, we investigate in section
\ref{sec:epcac} the
necessity and the consequences of a modification of the PCAC relation and
related current algebra results. The theoretical uncertainties and the
content of two assumptions introduced in section \ref{sec:epcac} is discussed
in
section \ref{sec:Content}.
In section \ref{sec:EmpB}, the empirical basis for our extension of PCAC
relation is discussed.
We find certain inconsistence that suggests the
need for a consideration of
diquark condensation  or an extension of the PCAC relation.
Section \ref{sec:Others} is devoted to other observations that are considered
to
favor the introduction of diquark condensation with different degrees
of certainty. We demonstrate the need for a further study of that
possibility with better theoretical and experimental
precision . Section \ref{sec:Sum} contains a summary.

\section{The chiral $SU(2)_L\times SU(2)_R$ Ward-Takahashi identity}
\label{sec:WT}

We study a chiral Ward-Takahashi identity in this section to show the
existence and the basic properties of the Goldstone diquark mode in
the superconducting phase. One way of studying the properties of pions
and Goldstone diquarks is to solve certain Bethe-Selpeter equation
provided that an explicit model, which is not important for the study of this
paper, is specified. The method used here depends upon
no particular model assumptions but the structure of the quark self
energy. In order to investigate the superconducting phase, an
8-component ``real'' representation for the quark field $\Psi$ is
introduced \cite{YING1}
\begin{eqnarray}
   \Psi &=&  \left ( \begin{array}{c} \psi \\ \tilde\psi \end{array}
              \right )\label{Psidef}
\end{eqnarray}
with $\psi$ and $\tilde\psi$ 4-component Dirac spinors. The ``real''
condition for $\Psi$ is
\begin{eqnarray}
   \bar\Psi &\equiv& \left ( \psi^\dagger \gamma^0, \tilde\psi^\dagger
   \gamma^0 \right ) = \Psi^T \Omega,\label{Psireal}
\end{eqnarray}
where
\begin{eqnarray}
  \Omega &\equiv& \left (\begin{array}{cc}
                        0     &     C^{-1}i\tau_2\\
                       C i\tau_2 & 0
                   \end{array} \right ),\label{Omegadef}
\end{eqnarray}
with superscript ``T'' denoting transpose,
$C$ the charge conjugation operator and $\tau_2$ the second Pauli
matrices acting on the flavor components of $\Psi$. The 8-component
spinor $\Psi$ for quarks given by Eq. \ref{Psidef} is used through out the
paper.

The axial vector current vertex $iA^{5a}_\mu(p',p)$ between single
quark states is written as
\begin{eqnarray}
 iA^{5a}_\mu(p+{q\over 2},p-{q\over 2}) &=& {i\over 4}
\gamma_\mu\gamma^5\tau^a O_3 + \Gamma^{5a}_\mu(p+{q\over 2},p-{q\over
2}),\label{WT1}
\end{eqnarray}
where $\Gamma^{5q}_\mu$ is the radiative part of $A^{5a}_\mu$,
the initial and final state (8-component) quark spinors are
suppressed and $q_\mu$ stands for the 4-momentum transfer. Here $O_3$
and $O_{(\pm)}$ in the following are Pauli matrices acting on the
upper and lower 4-components of the 8-component spinor
$\Psi$ \cite{YING1}.
$\Gamma^{5a}_\mu$ satisfies the chiral
Ward-Takahashi identity
\begin{eqnarray}
q^\mu\Gamma^{5a}_\mu(p+{q\over 2},p-{q\over 2}) = -{i\over 4}\left (
\Sigma\gamma^5\tau^a O_3+\gamma^5\tau^a O_3\Sigma\right ),\label{WT2}
\end{eqnarray}
where $\Sigma = \sigma - \gamma\cdot\phi^c\gamma^5 {\cal A}_c O_{(+)} +
\gamma\cdot \bar\phi_c\gamma^5 {\cal A}^c O_{(-)}$ is the self-energy
term (without the contribution to the wave function renormalization)
for the quarks and ${\cal A}_{ab}^c = - {\cal A}_{c,ab} = -\epsilon^{abc}$.
$\epsilon^{abc}$ is the total antisymmetric Levi-Civita tensor in the
color space of the quark. In both the Nambu phase where $\sigma \neq 0$
and $\phi^c_\mu = \bar\phi_{c\mu} = 0$ and the superconducting phase
where $\phi^c_\mu\neq 0$, $\bar\phi_{c\mu}\neq 0$ and possibly $\sigma
\neq 0$, Eq. \ref{WT2} implies that $\Gamma^{5a}_\mu$ contains a
massless Goldstone boson pole due to the fact that its right hand side
(r.h.s.) is finite in the $q^2\to 0$ limit. The appearance of this
massless pole in the physical excitation spectrum following the
spontaneous chiral symmetry breaking is required  by the Goldstone theorem.

 The chiral Ward-Takahashi identity Eq. \ref{WT2} can determine various
properties of the chiral Goldstone boson. We shall consider the case
in which  $\phi^2 \equiv \bar \phi_{c\mu}\phi^{c\mu} \neq 0$,
$\mu^\alpha = 0$ and $\sigma = 0$ that has not been discussed in the
literature to demonstrate some of the elementary features of the
superconducting phase related to the chiral symmetry. Eq. \ref{WT2}
becomes
\begin{eqnarray}
q^\mu\Gamma^{5a}_\mu(p+{q\over 2},p-{q\over 2}) &=& - {i\over 2} \left
( \gamma\cdot \bar\phi_c {\cal A}^c O_{(-)} + \gamma\cdot\phi^c{\cal
A}_c O_{(+)}\right ) \tau^a\label{WT3}
\end{eqnarray}
in such a case. The propagator of the Goldstone diquark is defined as
$G_\delta(q) = -i t^{\mu\nu}/\Delta(q)$. The denominator $\Delta(q)$
can be generally parameterized as
\begin{eqnarray}
\Delta(q) = q^2 + a_\delta{(\phi\cdot q)^2\over\phi^2}\label{DELTA}
\end{eqnarray}
if we choose the phases of $\phi^c_\mu$ and
$\bar\phi_{c\mu}$ such that $\phi^c_\mu = - \bar\phi_{c\mu}$. With the
following ansatz, namely,
\begin{eqnarray}
t^{c'\mu\nu}_c(q\to 0) = -{{\phi^{c'\mu}\bar\phi^\nu_c}\over{\phi^2}},
&&
\bar t^{\mu c\nu}_{c'}(q\to 0) = -{\bar
\phi{^\mu}_{c'}\phi^{c\nu}\over{\phi^2}},
\label{PROJ1}
\end{eqnarray}
and with the definition of the Goldstone diquark-quark vertices given
by
\begin{eqnarray}
\bar D^{ca}_\mu(p+{q\over 2},p-{q \over 2}) = - {i\over 2} g_{\delta
  q}\gamma_\mu \tau^a {\cal  A}^c O_{(-)}, &&
D^{a}_{c\mu}(p+{q\over 2},p-{q\over 2}) = {i\over 2} g_{\delta q}
\gamma_\mu \tau^a {\cal A}_c O_{(+)},\label{DS1}
\end{eqnarray}
we can obtain the value of $a_\delta$, the Goldstone diquark-quark coupling
constant $g_{\delta
  q}$ and the Goldstone diquark decay constant $f_\delta$ by
separating out the massless pole in
$\Gamma^{5a}_\mu$, which can be approximately evaluated by using an
one loop perturbation calculation \cite{YING2}. The result is
\begin{eqnarray}
\left (\Gamma^{5a}_\mu \right )_{pole} = -2 D^b_a\left ( p+{q\over 2},p-{q
\over 2}\right ) {-it^{\alpha\beta}\over {\Delta(q)}} Tr\int {d^4k\over
(2\pi)^4} \bar D^b_\beta\left (k-{q\over 2},k+{q\over 2}\right
)\nonumber \\
{i\over \gamma\cdot \left (k+{q\over 2}\right )-\Sigma}{i\over
2}\gamma_\mu\gamma^5{\tau^a\over 2} O_3 {i\over \gamma\cdot \left
(k-{q\over 2}\right)-\Sigma} - H.c.,\label{WTpole}
\end{eqnarray}
where $H.c.$ stands for the hermitian conjugation, $Tr$ stands for the
trace operation in the Dirac, flavor, color and upper and lower
4-component spaces of $\Psi$ and the color indices are suppressed. It
can be noticed that the symmetry factor \cite{YING2} for the Feynman
diagram is different from the 4-component theory for fermions. After
a lengthy process of evaluating the trace and performing the
4-momentum integration, the above equation together with Eq. \ref{WT3}
determine the values of
$a_\delta$, $g_{\delta q}$ and $f_\delta$ \cite{YING2} as functions of
$\phi^2$. The numerical values for them are shown in Figs.
\ref{Fig1}-\ref{Fig3}. A
Goldberger-Treiman relation for single quarks in the superconducting
phase exist; it can be expressed as $g_{\delta q} f_{\delta} =
\sqrt{\phi^2}$, which also defines the scale of $f_\delta$.

 The existence of the chiral Goldstone diquark in the superconducting
phase is thus established.

\section{Small current quark mass perturbation}
\label{sec:m0pert}

A finite current quark mass that explicitly breaks the chiral $SU(2)_L\times
SU(2)_R$ symmetry has non-trivial physical consequences.
For simplicity, we shall assume that both the up and down
current quarks have an identical mass $m_0$. Some of the consequences of a
finite mass for light quarks can be  studied  base on the Ward-Takahashi
identity given by Eq. \ref{WT2} with the term corresponding to the
divergence of the axial vector current operator taken into account, namely,
\begin{eqnarray}
q^\mu \Gamma^{5a}_\mu\left (p+{q\over 2},p-{q\over 2} \right ) &=&
{i\over 2} m_0 D_\pi \gamma^5\tau^a O_3 - {i\over 4}\left (
\Sigma\gamma^5\tau^a O_3 + \gamma^5\tau^a O_3\Sigma  \right ),
\label{WTmass1}
\end{eqnarray}
and
\begin{eqnarray}
{i\over 4}D_\pi \gamma^5 \tau^a O_3 &=& \int d^4x_1 d^4x_2
e^{ix_1\cdot(p+{q/2})-ix_2\cdot (p-q/2)}\mathopen{\langle 0\, |}
T\Psi(x_1)\bar\Psi(x_2) j^{5a}(0)\mathclose{|\,
0\rangle}|_{amp}.\label{WTmass2}
\end{eqnarray}
Here ``T'' stands for the time ordering, $j^{5a}= {i\over
4}\bar\Psi\gamma^5\tau^a O_3\Psi$ and the subscript ``amp'' denotes
the amputation of external fermion lines. $D_\pi$ is a scalar
function.

In the Nambu phase,  if the assumption that $D_\pi(q^2=0)$ is  dominated by the
pion pole, the mass of the pion moves  to
a finite value,  provided that to the first order in $m_0$, the
Goldberger-Treiman relation $g_{\pi q} f_\pi  = \sigma$ and the
Gell-Mann, Oakes and Renner \cite{GOR} (GOR) relation $f^2_\pi m^2_\pi
=-{1\over 2}m_0\mathopen{\langle 0\,|} \bar \Psi \Psi \mathclose{|\, 0\rangle}$
hold. This can be
checked by evaluating the r.h.s. of Eq. \ref{WTmass1}. The $\Sigma$
term on the r.h.s. of Eq. \ref{WTmass1} has a simple diagonal form
\begin{eqnarray}
\Sigma &=& \left (\begin{array}{cc}
                   \sigma & 0\\
                    0 & \sigma
                  \end{array}
           \right )\label{SigMat}
\end{eqnarray}
in the Nambu phase. It can be shown that Eq. \ref{WTmass1} is \cite{TEXTB}
\begin{eqnarray}
q^\mu \Gamma^{5a}_\mu\left (p+{q\over 2},p-{q\over 2} \right ) &=&
{i\over 2}\left ({m_0\over 2} {g_{\pi q}\over f_\pi\sigma} G_\pi(q^2)
\mathopen{\langle 0\,|} \bar\Psi\Psi\mathclose{|\, 0\rangle}-1\right
)\gamma^5\tau^a O_3,\label{NambuAdiv}
\end{eqnarray}
where
\begin{eqnarray}
G_\pi(q^2) = {1\over q^2-m_\pi^2} + \bar R(q^2).\label{Dpi}
\end{eqnarray}
$\bar R(q^2)$ is a smooth function of $q^2$ at small $q^2$ (namely,
$q^2 \leq m_\pi^2$). If the above mentioned Goldberger-Treiman
relation and the GOR relation hold and $\bar R(q^2=0) = 0$,
the r.h.s. of Eq. \ref{NambuAdiv} vanishes at $q^2=0$. Therefore
$\Gamma^{5a}_\mu$ is regular at $q^2 = 0$. It implies the
disappearance of the massless pole in $\Gamma^{5a}_\mu$ as well as in
the physical spectra.

In the superconducting phase where $\phi^2\neq 0$ and possibly
$\sigma\neq 0$, the situation is more complicated. In this case,
$\Sigma$ term on the r.h.s. of Eq. \ref{WTmass1} takes the following
form
\begin{eqnarray}
\Sigma &=& \left (\begin{array}{cc}
            \sigma & - \gamma\cdot \phi^c\gamma^5 {\cal A}_c\\
            \gamma\cdot\bar\phi_c\gamma^5 {\cal A}^c & \sigma
                  \end{array}
           \right ).
\end{eqnarray}
A finite $m_0$ for
the current quarks in this case does not render the r.h.s. of Eq.
\ref{WTmass1} vanish when $q^2\to 0$. There always remains a finite
strength of the massless excitation in $\Gamma^{5a}_\mu$ as long as
the mass  term is of the form ${1\over 2} m_0 \bar\Psi\Psi$. As a
consequence, there are massless excitations in the superconducting
phase even if the chiral $SU(2)_L\times SU(2)_R$ symmetry is
explicitly broken by $m_0$. There is no GOR type of relation for the
Goldstone diquark in the superconducting phase. In addition, certain
mixing between two sets of the auxiliary fields $\pi^a$ and
($\delta^{c a}_\mu,\bar\delta^a_{c\mu}$) \cite{YING1} is needed to
represent the Goldstone boson excitation in the superconducting phase
even when $m_0\neq 0$ and $\sigma = 0$. The above mentioned mixing provides
us with one of the motivations for the following extension of the PCAC
relation.

\section{The extension of the PCAC relation and current algebra results}
\label{sec:epcac}

The present day large scale strong interaction vacuum is expected to
be in the Nambu phase. There are a few obvious reasons for this
statement. First, the overwhelming color confinement at the present
day condition prevents the large scale superconducting phase in the
strong interaction vacuum scenario from been acceptable. Second, the
long range strong interaction force in the superconducting phase due
to the massless Goldstone diquark excitation inside the hadronic
system is absent in the experimental observations. However, localized
superconducting phases inside a baryonic system are not implausible
possibilities.

Due to the above considerations and the ones given in the introduction,
we explore in this section a subset of the
phenomenological consequences of the possibility in which the interior
of a nucleon contains diquark condensation that spontaneously breaks
the chiral $SU(2)_L\times SU(2)_R$ symmetry down to an flavor (isospin)
$SU(2)_V$ symmetry. The study of Ref. \cite{YING3} indicates that the
model Lagrangian introduced in Ref. \cite{YING1} indeed support such a
scenario when the coupling constant $\alpha_3$ is sufficiently large.

The on shell matrix elements of the axial vector current operator
between single  nucleon states can be parameterized as
\begin{eqnarray}
\mathopen{\langle p'\,|} A^a_\mu(0)\mathclose{|\, p\rangle} &=& \bar U(p')\left
(g_A\gamma_\mu+g_P
q_\mu + g_T{i\sigma_{\mu\nu}q^\nu\over 2 m_N}\right )\gamma^5
{\tau^a\over 2} U(p), \label{AxialMatr1}
\end{eqnarray}
with $q_\mu = (p'-p)_\mu$, $m_N$ the mass of a nucleon and U(p) the
4-component nucleon spinor. The longitudinal piece $g_P q_\mu$ on the
r.h.s. of Eq. \ref{AxialMatr1} is dominated by the contributions of the
Goldstone bosons of the spontaneous chiral symmetry breaking. If only
the Nambu phase is considered, $g_P$ is given by
\begin{eqnarray}
g_P(q^2) &=& - 2 {g_{\pi NN}(q^2)f_\pi(q^2)\over {q^2-m_\pi^2}} + \bar
g_P(q^2),\label{GP}
\end{eqnarray}
with $\bar g_P(q^2)$ the residue term  and $g_{\pi NN}(q^2)f_\pi(q^2)$
a slow varying function of both $m_\pi^2$ and $q^2$. If, however, the
assumption that there is a diquark condensation inside a nucleon is
made, there world be another longitudinal term in the matrix elements
of the axial vector current operator due to the Goldstone diquark
excitation inside  that nucleon. The expression for $g_P$ has to be
modified to
\begin{eqnarray}
g_P(q^2) &=& -2 \left ({g_{\pi NN}(q^2)f_\pi(q^2)\over {q^2-m_\pi^2}} +
                      z_\delta g_{\delta
N}(q^2)f_\delta(q^2)\eta(q^2)\right ) + \bar g_P(q^2)\label{GP2}
\end{eqnarray}
after considering this additional excitation. Here $\eta(q^2)$ is
related to the propagator of the Goldstone diquark excitation and
$z_\delta$ is a constant. Similar to the pion, we introduce $g_{\delta N}$
as the Goldstone diquark-nucleon coupling constant and $f_\delta$ as
the Goldstone diquark decay constant. Albeit there is massless
excitation in $\Gamma^{5a}_\mu$, there is  no pole behavior in
$\eta(q^2)$ in  the small $q^2$ region due to the fact that a
Goldstone diquark  carries color so that it, like a quark, is
confined inside the nucleon.

At the operator level, the divergence of the axial vector current
operator is
\begin{eqnarray}
     \partial^\mu A^a_\mu &=& {1\over 2}m_0
     \bar\Psi i\gamma^5\tau^a O_3\Psi.\label{DA5}
\end{eqnarray}
The PCAC relation is given by
\begin{eqnarray}
\partial^\mu A^a_\mu &=& - f_\pi m_\pi^2 \phi_\pi^a,\label{PCAC2}
\end{eqnarray}
which makes an implicit assumption that the quark bilinear
operator $\bar\Psi i\gamma^5\tau^a O_3\Psi$ couples only to the pion
excitation in the low momentum transfer regime.
It can be regarded as a definition of the pion field
(when going off the pion mass shell). Taking the matrix elements
of Eq. \ref{PCAC2}, it can be shown that this definition is
inconsistent with Eq. \ref{GP2} due to the additional term added to
$g_P$. We have at least two choices. The first one is to reject
Eq. \ref{GP2}, which we shall not do in this paper. The second one is
to modify Eq. \ref{PCAC2} when its matrix elements are taken.
We take the following assumption that
\begin{quote}
{\bf Assumption I:} {\em The divergence of the axial vector current
  operator is dominated by the longitudinal chiral Goldstone bosons, namely,
the
  pion and the possible Goldstone diquark, contributions in
  the low momentum transfer region.}
\end{quote}
in the sequel. The content of it, especially the range of $q^2$ in
which it is valid, will be explained in the next section
in more detail

      For a nucleon, it can be specified as
\begin{eqnarray}
  \mathopen{\langle p'\,|} \partial^\mu A^a_\mu \mathclose{|\, p\rangle} &=& -
\mathopen{\langle p'\,|}\left (
f_\pi m_\pi^2 \phi^a_\pi + f_\delta s_\delta \phi_\delta^a\right )
\mathclose{|\, p\rangle},
\label{PCAC3}
\end{eqnarray}
with $s_\delta \sim m_\pi^2$ a parameter proportional to $m_0$ that
characterizes the strength of mixing of the Goldstone diquark
excitation within the $\bar\Psi i\gamma^5\tau^a O_3\Psi$ operator in the
presence of baryonic matter and
$\phi_\delta^a$ a pseudo-scalar \cite{drive}
driving field for the Goldstone diquark excitation inside the nucleon.
The physical meaning of Eq. \ref{PCAC3} is that in a baryonic system,
the operator $\partial^\mu A^a_\mu= {1\over 2}
m_0 \bar\Psi i \gamma^5 \tau^a O_3
\Psi$  can excite  two sets of distinct longitudinal chiral soft modes
(Goldstone bosons in the chiral symmetric limit) if there is diquark
condensation.

{}From Eqs. \ref{AxialMatr1} and \ref{PCAC3}, one obtains the following
equation
\begin{eqnarray}
0 &=& q^2 \left [2 m_N g_A(q^2) +
  (q^2-m^2_\pi)g_P(q^2)\right ]\nonumber\\
  &&\hspace{0.1in} + 2 m^2_\pi \left [ g_{\pi NN}(q^2)f_\pi(q^2)+(q^2-m^2_\pi)
      {s_\delta\over m^2_\pi}g_{\delta
        N}(q^2)f_\delta(q^2)\eta(q^2)-m_N g_A(q^2)\right ]\label{MAdiv}.
\end{eqnarray}
If the assumption that
\begin{quote}
{\bf Assumption II:} {\em $2 m_N g_A(q^2) + (q^2-m^2_\pi) g_P(q^2)$
  is a slow varying function of $q^2$ and $m^2_\pi$}
\end{quote}
is made, two equations follow, namely,
\begin{eqnarray}
g_P(q^2) &=& -2 {m_N g_A(q^2)\over q^2 - m_\pi^2},\label{PCAC4}\\
 m_N g_A(q^2) &=&
g_{\pi NN}(q^2) f_\pi(q^2) + (q^2 - m_\pi^2) {s_\delta \over m_\pi^2}
g_{\delta N}(q^2) f_\delta (q^2) \eta (q^2). \label{PCAC5}
\end{eqnarray}
The slow
varying assumption of these quantities together with the modified PCAC
relation, Eq. \ref{PCAC3}, imply that the residue term $\bar g_P(q^2)$
in Eq. \ref{GP2} is unimportant in the small $q^2$ regime, which is
consistent with the assumption I. Eqs.
\ref{PCAC4}, \ref{PCAC5} are consistent with Eq. \ref{GP2} provide that
$z_\delta =
s_\delta/m_\pi^2$. The modified  Goldberger-Treiman relation is
obtained when $q^2 = m_\pi^2$ is  assumed on the r.h.s. of the second
equation of Eq. \ref{PCAC5} and $q^2\to 0$ is taken on its left hand
side (l.h.s.), namely,
\begin{eqnarray}
m_N g_A(0) &=& g_{\pi NN}(m_\pi^2)f_\pi(m_\pi^2) + \lim_{q^2\to m_\pi^2}
(q^2-m_\pi^2)z_\delta g_{\delta N}(q^2) f_\delta(q^2)\eta(q^2).\label{GP3}
\end{eqnarray}
Assuming the results of Ref. \cite{GUERRA} can be used in the time like
region for $q_\mu$, the extrapolation of $g_A$ from $q^2 = m_\pi^2$ to
$q^2=0$ introduces an error of order $m_\pi^2/M_A^2\sim 1 \%$ with $M_A
\sim 1 GeV$ \cite{GUERRA}. Since there is no pole in $q^2$ for
$\eta(q^2)$ at $m_\pi^2$, Eq. \ref{GP3} reduces to
\begin{eqnarray}
  m_N g_A(0) &=& g_{\pi NN}(m_\pi^2)f_\pi(m_\pi^2),\label{GP4}
\end{eqnarray}
namely, the Goldberger-Treiman relation in a different form from the
one presented in the literature, which is
\begin{eqnarray}
  m_N g_A(0) &=& g_{\pi NN}(0)f_\pi(0).\label{GP5}
\end{eqnarray}

  Another form of the Goldberger-Treiman relation, which is called
the on shell Goldberger-Treiman relation, is important for the
discussions to be presented. It is
\begin{eqnarray}
  m_N g_A(m_\pi^2) &=& g_{\pi NN}(m_\pi^2)f_\pi(m_\pi^2).\label{GP6}
\end{eqnarray}

Eqs. \ref{PCAC4} and \ref{PCAC5} are the main results following
assumptions I and II.

The modification of some of the current algebra results  related to
PCAC due to a possible Goldstone diquark excitation inside a nucleon
can also be investigated. The Adler-Weisberger sum rule    and the
nucleon $\Sigma_N$ term will be studied in this paper base on a Ward
identity involving the axial vector current operator \cite{ADLER,WEISB,BPP}.
In order to obtain useful  information, the low lying longitudinal excitation
contributions and the rest part of the axial vector current operator
inside a time ordered product are separated in the following way
\begin{eqnarray}
\left <T\left (\ldots A_\mu^a \ldots \right )\right > &=& \left <
T\left (\ldots\bar A_\mu^a\ldots\right )\right > +\nonumber\\
&& \partial_\mu
\left < T\left (\ldots f_\pi\phi_\pi^a\ldots \right )\right > +
\partial_\mu\left < T\left (\ldots z_\delta
f_\delta\phi_\delta^a\ldots\right )\right >, \label{Adecomp}
\end{eqnarray}
with the second and the third terms on the r.h.s. the longitudinal
parts of $A_\mu^a$, which is dominated by the low lying chiral
Goldstone boson contributions, and $\bar A_\mu^a$, which is expected
to change slowly with the momentum transfer $q^\mu$ when its matrix
elements are taken between nucleon states, containing the rest part of
$A_\mu^a$. The matrix elements are evaluated on the pion mass shell
\cite{BPP}. As a rule, which is going to be explored in more detail in
other work \cite{ExtrapRule}, those of $\bar A_\mu^a$ that connect
nucleon state to other hadronic intermediate states  by
a gradient-coupling \cite{Weinberg} are
then extrapolated to the kinematic point where $q^2=0$ in order to
compare with experimental data by taking the assumption that
\begin{quote} {\bf Assumption III:} {\em they are
slow varying function of $q^2$}.
\end{quote}
The  error of the extrapolation is
expected to be of order $O(m_\pi^2/M_A^2)\sim 1\%$. There are
contributions from other off (nucleon) shell terms and baryonic
excitations in the intermediate states if $q^2\ne 0$;
At finite $q^2$, the contribution  from the single nucleon pole can
not be  separated from the background. The extrapolation to $q^2=0$
allows the extraction of single nucleon axial vector form factor $g_A(0)$.
The modified Adler-Weisberger sum rule can be shown to
have the following form
\begin{eqnarray}
g_A^2(q^2=0)&=&1-2{f_\pi^2(m_\pi^2)\over \pi}\int_{m_\pi}^\infty d\nu
{\sigma^{\pi^-p}_{tot} (\nu)-\sigma^{\pi^+ p}_{tot}(\nu)\over
  (\nu^2-m_\pi^2)^{1/2}}
\nonumber\\
&& - 2\lim_{q^2\to m_\pi^2}(q^2-m_\pi^2)^2 f_\delta^2(q^2)\lim_{\nu
  \to 0} {z_\delta^2\over\nu} G_{\delta N}^{(-)}(\nu,0,q^2,q^2),\label{gA2}
\end{eqnarray}
where $G_{\delta N}^{(-)}$ is related to the flavor odd forward
Goldstone diquark-nucleon scattering amplitude, which is driven by
$\phi_\delta^a$, without the amputation of the diquark lines.
Assumption III is nontrivial. It is, however, supported by
observations. This is discussed in section \ref{sec:EmpB}.

The expression for the nucleon $\Sigma_N$ term, which is of order
$m_0$, is modified to
\begin{eqnarray}
\Sigma_N &=& f_\pi^2(m_\pi^2) T^{(+)}_{\pi NN}(0,0,m_\pi^2,m_\pi^2) +
\lim_{q^2\to m_\pi^2} (q^2-m_\pi^2)^2 f_\delta^2(q^2)z_\delta^2 G_{\delta
  N}^{(+)}(0,0,q^2,q^2),\label{SigmaN}
\end{eqnarray}
with $G^{(+)}_{\delta N}$ related to the flavor even forward
Goldstone diquark-nucleon scattering amplitude without the amputation
of the diquark lines. The second term on the r.h.s. of Eq. \ref{SigmaN}
vanishes on the pion mass shell.

\section{The content of assumptions I and II and theoretical
  uncertainties}
\label{sec:Content}

  Before confront the results of the previous sections with
observations, we assess the possible theoretical uncertainties
involving assumptions I and II to ascertain the degree of confidence
that one should assume for the discussions in the following sections.

Assumptions I and II are sufficient conditions for Eqs.
\ref{PCAC4} and \ref{PCAC5}. They are, however, not necessary ones for
Eq. \ref{PCAC4}. This is because Eq. \ref{PCAC4} represents the
conservation of axial vector current in the chiral limit of $m_0\to 0$
or in the high momentum transfer limit of $q^2 >> m_\pi^2$.
The true non-trivial $q^2$ region for Eq. \ref{PCAC4} consists of low
$q^2$, where the axial vector current is not conserved. The $q^2$
region in which Eq. \ref{PCAC5} hold contains the $q^2$ region
where assumptions I and II are true. It can be parameterize as
$ -\zeta_- s_{th} <
q^2 < \zeta_+ s_{th}$ with $s_{th} \sim 1$ $\mbox{GeV}^2$ and
the values for $\zeta_-$ and $\zeta_+$ determined in the
following. The above statements can be justified through a more detailed
analysis of the
content of assumption II by relaxing the constraint imposed by
assumption I. To this end, Eq. \ref{MAdiv} is rewritten as
\begin{eqnarray}
   q^2 A(q^2,m_\pi^2) +  m_\pi^2 B(q^2,m_\pi^2) &=& m_\pi^2
   C(q^2,m_\pi^2), \label{MAdiv1}
\end{eqnarray}
with
\begin{eqnarray}
   A(q^2,m_\pi^2) &=& m_N g_A(q^2) + {1\over 2}
                   (q^2-m_\pi^2) g_P(q^2), \label{Adef}\\
   B(q^2,m_\pi^2) &=& g_{\pi NN}(q^2)f_\pi(q^2)+(q^2-m^2_\pi)
      {s_\delta\over m^2_\pi}g_{\delta
        N}(q^2)f_\delta(q^2)\eta(q^2)-m_N g_A(q^2), \label{Bdef}
\end{eqnarray}
and $C(q^2,m_\pi^2)$ to be defined in the following.
$C(q^2,m_\pi^2)=0$ when assumption I is valid. Here the $m_\pi^2$
dependence of $A$, $B$ and $C$ is written out explicitly.
The following equations hold, namely,
\begin{eqnarray}
   \lim_{q^2\to \infty} A(q^2,m_\pi^2) &=& 0, \label{Aconsv1} \\
   \lim_{m_0\to 0} A(q^2,m_\pi^2) & = & 0, \label{Aconsv2}
\end{eqnarray}
which represents the conservation of axial vector current in the
$q^2 >> m_\pi^2$ region and in the chiral symmetric limit ($m_0\to
0$). Assumption
II is non-trivial since it implies $A(q^2,m_\pi^2) \approx
\lim_{m_0\to 0} A(q^2, m_\pi^2) = 0$, which establishes Eq.
\ref{PCAC4} for any $q^2$.

If $A(q^2,m_\pi^2) =
0$, then Eq. \ref{MAdiv1} reduces to
\begin{eqnarray}
   B(q^2,m_\pi^2) &=& C(q^2,m_\pi^2). \label{MAdiv2}
\end{eqnarray}
That $C(q^2,m_\pi^2)$ is small is the true content of assumption I.
This can be seen from the following analysis.

Let's define an error operator for our extended PCAC relation,
\begin{eqnarray}
  m_\pi^2 \hat\Delta^a(x) & = & \partial^\mu A^a_\mu(x) - m_0
  \chi^a(x),\label{ErrPCAC1}
\end{eqnarray}
with the longitudinal chiral soft excitation modes driving field
$\chi^a(x)$
\begin{eqnarray}
    \chi^a(x) &=& {1\over 2} P(s_{th}) \bar \Psi(x) i\gamma^5 \tau^a O_3\Psi(x)
                  P(s_{th}),\label{ChiDef}
\end{eqnarray}
where the projection operator $P(s_{th})$ is defined as
\begin{eqnarray}
    P(s_{th})\mathclose{|\, s\rangle} &=& 0
\end{eqnarray}
for any state with pionic quantum number and invariant mass $s>s_{th}$ and
\begin{eqnarray}
    P(s_{th})\mathclose{|\, s\rangle} &=& \mathclose{|\, s\rangle}
\end{eqnarray}
for any state with pionic quantum number and invariant mass $s\le
s_{th}$. The value for $s_{th}$ is determined in the following.
The matrix element of $\hat \Delta^a(0)$ between
single nucleon states is defined as
\begin{eqnarray}
  C(q^2) \bar U(p') i\gamma^5 \tau^a U(p) &=& \mathopen{\langle \,
    p'|}
   \hat \Delta^a(0)\mathclose{|\, p\rangle},\label{Cdef}
\end{eqnarray}
where $C(q^2)$ is identical to the $C(q^2,m_\pi^2)$ in Eq.
\ref{MAdiv1}.
An once subtracted dispersion relation for $C(q^2)$ can be used, namely,
\begin{eqnarray}
  C(q^2) &=& C(m_\pi^2) + {q^2-m_\pi^2\over \pi} \int^\infty_{s_{th}}
   ds'  {Im C(s')\over (s'-q^2)(s'-m_\pi^2)}. \label{Cdispers}
\end{eqnarray}

At low $q^2$, the value of the integral on the r.h.s. of the above
equation is insensitive to the shape of $ImC(s)$ as a function of $s$,
so it can be replaced by a step function of the following form
\begin{eqnarray}
   ImC(s) = \alpha \theta(s-s_{th})
\end{eqnarray}
for our purpose,
with
\begin{eqnarray}
\alpha \sim  C(m_\pi^2). \label{alphaC}
\end{eqnarray}
The above relation is a consequence of the property that in an asymptotic
free theory like QCD, the form factor $C(q^2)$ actually satisfies an
unsubtracted dispersion relation
due to the fact that $ImC(q^2)$ falls off faster
than a constant at large momentum transfer squared $s$, where the
dominant contribution to the matrix element of
operator $\hat\Delta(0)$ between initial vacuum state
and hadronic final states is through its coupling to the current
quark-antiquark lines. The once subtracted form of dispersion
relation is used to emphasize the small value of $C(m_\pi^2)$ in
observations.

This allows the integration in Eq. \ref{Cdispers}
to be done explicitly
\begin{eqnarray}
 C(q^2) = C(m_\pi^2) + {\alpha\over \pi} ln\left ( {
s_{th} - q^2 \over s_{th} - m_\pi^2 }\right ). \label{Cq2dep}
\end{eqnarray}
In general $C(q^2)$ is not necessarily small at low $q^2$ since the term
corresponding to it is of order $m_\pi^2$ already. Theoretically,
the smallness of this
quantity at low $q^2$ is a non-trivial result which can be argued
by using the asymptotic freedom property of QCD.
The fact $C(m_\pi^2)$ is small can be deduced from the success of the
Goldberger-Treiman relation on the pion mass shell
(without extrapolation in $q^2$) given by
Eq. \ref{GP6} in the
empirical observations\cite{GTDisp}. The error for it has an order of
magnitude
of around $1\%$. The range of $q^2$ in which assumption I is valid
can be obtained by requiring $\delta C(q^2) \equiv |C(q^2)-C(m_\pi^2)|$
to be less than or equal to $|C(m_\pi^2)|$.
This gives $q^2 < 0.7s_{th}$.
The range of validity of the above arguements in the negative
$q^2$ region is much larger than 1 $\mbox{GeV}^2$ in magnitude.

The next step is to estimate the value of $s_{th}$. The lowest
value of $s_{th}$ in the pionic channel is below $9 m_\pi^2$, which
corresponds to an
anomalous threshold for the three pion state. However, the
effective $s_{th}$ correspond to that of the $\rho\pi$ threshold,
which is considerablly larger than $9 m_\pi^2$. This is a consequence of
the fact that the underlying dynamics of QCD is chiral invariant
except for a small mass term. From this fact the dynamical (operator)
equation Eq. \ref{DA5} follows. This dynamical equation ensures that
the operator $\bar\Psi i\gamma^5 \tau^a O_3\Psi$ can only excite a
longitudinal vector excitation since it is proportional to the
divergence of a axial vector operator field. Therefore the state in which
the three pions are all in a s-state is dynamically forbiden. The allowed
state which dominates the dispersion relation is the one where
two of the three pions have a relative angular momentum  of 1, which
is itself dominated by the $\rho$ excitation strength. Therefore $s_{th}\sim
(m_\rho+m_\pi)^2$ and the range of $q^2$ in which assumption I
and II is valid is 30 to 35 times larger than $m_\pi^2 \approx
0.02 GeV^2$.

   In conclusion, both of assumptions I and II are supported by
observations and the physical picture based on the smallness of the
current quark masses. Assumption I is connected to the success of the
on pion mass shell (without $q^2$ extrapolation) Goldberger-Treiman
relation given by \ref{GP6}; assumption II is connected to the
verification of Eq. \ref{PCAC4} in an experimental investigation
 \cite{choi}.

\section{Empirical basis}
\label{sec:EmpB}

\subsection{Goldberger-Treiman relation}

The Goldberger-Treiman relation is satisfied to about $5\%$ in
experimental observations. This good agreement indicates  the validity
of the various assumptions combined made above and a rather small
value of $\lim_{q^2\to m_\pi^2} (q^2-m_\pi^2)z_\delta g_{\delta
N}(q^2)f_\delta(q^2)\eta(q^2)$. The $\eta(q^2)$ term on the pion mass
shell actually  vanishes since $\eta(q^2)$ does not has a pole in the
low $q^2$ region. However, a small value for the $\eta(q^2)$ term
does not  follow from the success  of the Goldberger-Treiman relation.

 Eq. \ref{GP5} is the other form of the Goldberger-Treiman relation
used in the literature. It can be made to contain an error of less
than $1\%$ if soft $g_{\pi NN}(q^2)$ is used \cite{GTDisp}. Eq.
\ref{GP6} is an equation that is satisfied in observation to an
accuracy of $1\%$. The accuracy of Eq. \ref{GP6} in
observations forms  one of the empirical basis for assumption I. This
point is discussed in some detail in section \ref{sec:Content}.

\subsection{Nucleon axial vector form factor $g_A$ and pseudoscalar
  form factor $g_P$}

A recent experimental study \cite{choi} provides a strong support
\cite{explain}
of Eq. \ref{PCAC4} within momentum transfer region of $0<-q^2 < 0.2\hspace{3pt}
GeV^2$.

\subsection{$g_A$ and the pion-nucleon coupling constant $g_{\pi
    NN}$}

Eq. \ref{PCAC5} gives a specific relation between $g_A(q^2)$ and
$g_{\pi NN}(q^2)f_\pi(q^2)$ if no other chiral soft modes bellow $q^2 =
(m_\rho+m_\pi)^2$  is present.
We study whether or not Eq. \ref{PCAC5} without the Goldstone
diquark contribution term, namely,
\begin{eqnarray}
    m_N g_A(q^2) &=& g_{\pi NN}(q^2) f_\pi(q^2) \label{PCAC6}
\end{eqnarray}
 is consistent with phenomenology.

  The $q^2$ dependence of $g_A(q^2)$ in the space-like $q^2$ region
is of a dipole form \cite{GUERRA}, namely,
\begin{eqnarray}
      g_A(q^2) &=& { {g_A(0)}\over {
   \left (1 - {q^2 / M_A^2} \right )^2}},\label{g_Aq2}
\end{eqnarray}
with $M_A \approx 1$ GeV. We shall use $M_A=1.0$ GeV in the following.

   The $q^2$ dependence of $g_{\pi NN}(q^2)$ is known less well than
that of $g_A(q^2)$. A monopole form for it, which can be parameterized
as
\begin{eqnarray}
      g_{\pi NN}(q^2) &=& g_{\pi NN}
        {{\Lambda_{\pi NN}^2 - m_\pi^2}
        \over
        {\Lambda_{\pi NN}^2 - q^2}} \label{gpiNNq2}
\end{eqnarray}
is agreed upon in the literature; the value for $\Lambda_{\pi NN}$
varies. It is found to be greater than 1.2 GeV in nucleon-nucleon
(NN) scattering and deuteron property studies \cite{Bonn}. This range
of $\Lambda_{\pi NN}$ implies a rather small quark core for a nucleon.
It is in contradiction with the expectations of many chiral nucleon
models for the nucleon (see, e.g. Refs. \cite{Cohen}--\cite{Thomas}).
Lattice QCD evaluation \cite{LatQCDg} also indicates
a smaller one, namely, $\Lambda_{\pi NN} \sim 800$ MeV. On the
phenomenological side, Goldberger-Treiman discrepancy study
\cite{GTDisp},
$pp\pi^0$ v.s. $pn\pi^+$ coupling constant difference \cite{pppi0},
high energy $pp$ scattering \cite{ppscat} and charge exchange reaction
\cite{q-exch}, etc, support a value of $\Lambda_{\pi NN}$ close to
$800$ MeV. By introducing a second ``pion'' $\pi'$ with mass 1.3
GeV, $\Lambda_{\pi NN}$ can be chosen to be around $800$ MeV
without spoiling the fit to the NN scattering phase shifts and
deuteron properties \cite{Holinde}. This picture was later
justified by a microscopic computation in Ref. \cite{Janssen}.

   The $q^2$ dependence of $f_\pi(q^2)$ is little known from
experimental observations. It can be expressed in terms of a once
subtracted dispersion relation, namely,
\begin{eqnarray}
    f_{\pi}(q^2) &=& f_\pi(m_\pi^2) + {q^2 - m_\pi^2 \over \pi}
   \int_{s_{th}}^\infty ds' {Im f_\pi (s')\over (s'-q^2)
     (s'-m_\pi^2)}.\label{fpidisp}
\end{eqnarray}
Since the lightest physical state connects to the axial vector current
operator with the quantum number of pion is
the $\rho\pi$ two particle state, the value of $s_{th}$ is chosen to
be $(m_\rho + m_\pi)^2$, where $m_\rho = 770$ MeV. At $s'= 4 m_N^2$,
which correspond to the lowest invariant mass of a $\bar N N$ system,
another branch cut for $f_\pi(q^2)$ develops.
We shall include $\rho\pi$ state only since $\bar N N$
state contributions to Eq. \ref{fpidisp} is small when $q^2$ is small.
So Eq. \ref{fpidisp} can be written as
\begin{eqnarray}
    f_{\pi}(q^2) &=& f_\pi(m_\pi^2) + {q^2 - m_\pi^2 \over \pi}
   \int_{s_{th}}^{4 m_N^2} ds' {Im f_\pi (s')\over (s'-q^2)
     (s'-m_\pi^2)}.\label{fpidisp1}
\end{eqnarray}
Our next step involves the specification or computation of $Im
f_\pi(s)$ by exploring the fact that only the $\rho\pi$ state which couples
to the pion contributes to $Im f_\pi(s)$ in the momentum transfer
region of interest to this paper. An one loop computation to $Im
f_\pi(s)$ is known to be insufficient to account for experimental data
in other studies \cite{Holinde,Janssen}, the $\rho \pi$ correlation,
which forms a resonance near $1.3$ GeV, is required. We therefore
propose the following form for $Im f_\pi(s)$,
\begin{eqnarray}
    Im f_\pi(q^2) &=& {3\over 4} {m_N\over g_{\pi NN}}
    {g_{\rho\pi\pi}^2\over 4 \pi}\bar\rho_{\pi,\rho\pi}(q^2)
    \label{Imfpirho}
\end{eqnarray}
with the reduced density of state
\begin{eqnarray}
    \bar\rho_{\pi,\rho\pi}(q^2) &=& \bar\rho_0(q^2) \left ( 1 + {\lambda I_B
      \over (q^2-s_B)^2 + \bar\rho^2_0(q^2) I_B^2}
     \right )\label{rhobar}
\end{eqnarray}
and
\begin{eqnarray}
   \bar \rho_0(q^2) &=& \sqrt{1-{(m_\rho+m_\pi)^2\over q^2}}
    \sqrt{1 - {(m_\rho-m_\pi)^2\over q2}}
   \left (1 -{m_\rho^2-m_\pi^2\over 3 q^2}   \right )
   \theta[q^2-(m_\rho+m_\pi)^2],\label{barrho0}
\end{eqnarray}
where $\theta(x)$ is the step function with a value of unity for
positive x,
$s_R$ is  chosen to be $1.69$ $\mbox{GeV}^2$, $\lambda$ characterizes
the strength of the $\rho \pi$ resonance in $Im f_\pi(q^2)$ and
$\bar\rho_0(s_R) I_B$ characterizes the width of the resonance.

The above form is chosen so that when $\lambda=0$, $Im f_\pi(q^2)$ is
the one loop result in the Feynman-t' Hooft gauge (for the $\rho$
propagator). The $\rho\pi\pi$ interaction piece of Lagrangian density
used for evaluation of $Imf_\pi(q^2)$ is
\begin{eqnarray}
      {\cal L}_{\rho\pi\pi} &=& g_{\rho\pi\pi}
          \epsilon^{abc} \pi^a \partial^\mu \pi^b \rho_\mu^c.\label{Lrhopipi}
\end{eqnarray}
The corresponding piece of the
axial vector current operator, which can be obtained
from the Noether theorem, can be written as
\begin{eqnarray}
    A^a_\mu = g_{\rho\pi\pi} {m_N\over g_{\pi NN}}
    \epsilon^{abc}\pi^b\rho_\mu^c,\label{Arhopipi}
\end{eqnarray}
where use has been made of the linear $\sigma$ model \cite{LinearSig}
relation $m_N = g_{\pi NN}\sigma$.

The value for $\lambda$, $\Lambda_{\pi NN}$ and $I_B$ is
adjusted so that the minimum value of the following function
\begin{eqnarray}
  f(\lambda,\Lambda_{\pi NN},I_B) &=& {1\over N}\sum_{k=0}^N
  {\left [m_N g_A(q^2_k)-g_{\pi NN}(q^2_k)f_\pi(q^2_k)   \right ]^2
   \over m_N^2 g^2_A(q^2_k) } e^{3 q^2_k},\label{minFunc}
\end{eqnarray}
with
\begin{eqnarray}
  q^2_k &=& q^2_{min} + k {q^2_{max}-q^2_{min}\over N},\label{q2k}
\end{eqnarray}
is achieved. The value of $N$ is chosen to be 100. $q_{min}^2 =
-0.6$ GeV and $q^2_{max} = 0.2$ GeV. The factor $e^{3q^2}$ is
used to put more weight on small $|q^2|$ region where the fit tends to
be poor.

In all cases listed in Table \ref{Table1}, $\Lambda_{\pi NN} < 0.95$ GeV. If a
value $g_{\rho \pi\pi}^2/4\pi = 1.0$ is taken, the qualitative shape
of $Imf_\pi(q^2)$ in Fig. \ref{Fig4} obtained by a minimization of Eq.
\ref{minFunc} is similar to the $Im\Gamma(q^2)$ of Ref.
\cite{Janssen}. Quantitatively, it has a broader width. The
phenomenological value for $g_{\rho\pi\pi}$ can be deduced from
the $\rho\to \pi \pi $ decay process. It has a value satisfies
$g_{\rho\pi\pi}^2/4\pi \approx 2.9$. Using this value of
$g_{\rho\pi\pi}$, $Imf_\pi(q^2)$ is obtained by minimization of
Eq. \ref{minFunc}. The result is given in Table \ref{Table1} and plotted in
Fig.
\ref{Fig4}. It's drastically different in shape from that of $Im\Gamma(q^2)$ in
Ref. \cite{Janssen}. In fact, instead of increasing the density of
states relative to the one loop result (Eq. \ref{Imfpirho} with
$\lambda=0$), the resonance
contribution decreases the density of states in order to satisfy
Eq. \ref{PCAC6}. The reduction of density of state indicates either the
solution is unphysical (for a normal resonance) or there is a
competing resonance in another channel that couples to the pionic
channel we are dealing with. But what can the ``other resonance''
channel be? There is no known resonance there. Therefore, the result
Eq. \ref{PCAC6} obtained from chiral symmetry argument without diquark
condensation for a nucleon is inconsistent with phenomenology. The
introduction of the Goldstone diquark contribution term seems to be
inevitable.

\subsection{Adler-Weisberger sum rule}

The value of $g_A$ from nuclear $\beta$ decay experiments is $1.261\pm
0.004$ \cite{PREV}. The value of the same quantity obtained from the
Adler-Weisberger sum rule using pion nucleon scattering cross section
(and  after estimating other corrections) is about $1.24\pm 0.02$
 \cite{ADLER}. These two numbers, with a central $g_A$ value
deviation of order $1.3\%$, provide a support for assumption III.
Again, the contribution of
the second term on the r.h.s. of Eq. \ref{gA2} is expected to be zero
on the pion mass shell. Therefore the good agreement of the
Adler-Weisberger sum rule with the observation does not necessarily
constitute a fact that is against the additional terms added in Eqs.
\ref{PCAC3} and \ref{Adecomp}.

\section{Other empirical observations}
\label{sec:Others}

Several  empirical observations, which are considered relevant to this
paper and for which the present understanding
for them in terms of conventional picture is considered to have
difficulties are listed and discussed in this section.
Since the previous theoretical understanding of the experimental
observations had not taken into account of the possibility of diquark
condensation inside a nucleon/nucleus, and, in addition,
some of the failure of the
conventional explanation for them is not unambiguous at the present
level of theoretical and/or experimental precision, they are
considered more as discussions than evidences. It serves the purpose
of either sharpening the problems, which is expected to lead to further
investigations, or showing
that the introduction of diquark condensation inside a nucleon/nucleus
does not contradict well established experimental facts.

\subsection{$g_P$ in a nucleus}

The effects of the Goldstone diquark inside a nucleon, if exist,
might be comparatively large in the kinematic region off the pion
mass shell. Current experimental data considered do not allow any
conclusive statement to be made on this point. The value of $g_P$
can be compared to the PCAC  value (Eq. \ref{GP}) to detect the
possible effects of the Goldstone diquark. Experimental determinations
of $g_P$ for a nucleon in muon capture experiments involving
light nuclei show a systematic increase (of order as large as $100\%$)
of the value of $g_P$  from its PCAC one \cite{TOWNER}. The world
average value of $g_P$ obtained from the muon capture experiments on
a hydrogen is close to the one given by the PCAC one \cite{BARDIN1}.
However, the interpretation of the results is not unambiguous
(Ref. \cite{BARDIN1} and Gmitro and Truol in Ref. \cite{TOWNER}).
There are two
recent measurements of the muon capture rate on the deuterium. The
central value of the first measurement \cite{BARDIN2} implies a value of
$g_P$ smaller than the PCAC one \cite{TATARA}. The central value of the
second measurement \cite{CARG} implies a value of $g_P$ close to the
PCAC one \cite{TATARA}. The problems related to the value of $g_P$ are not
yet completely settled \cite{RECENT}.
Polarized $\beta$ decay experiments offer
alternative means of measuring the value of $g_P$ in a different range
of $q^2$ \cite{YING4}. The experiments are difficult but in principle
possible.

Theoretically, the agreement of the PCAC value for $g_P$ and the
experimentally measured one is expected for a nucleon in free space.
This is because Eq. \ref{PCAC4} implies the deviation of $g_P$ from the
PCAC one is related to the deviation of the value of $g_A$, $m_N$ and $m_\pi$
from the experimentally observed ones, which is not true. Therefore,
the system in which a deviation of the strength of the pionic longitudinal
modes in the axial vector current operator from the PCAC
one can be observed is inside a nucleus with relatively large number
of nucleons. In these systems, the coupling between the pions outside
of the nucleons and the Goldstone diquark excitations inside the
nucleons could render the properties of pions to change to such a degree that
a deviation from PCAC value can be observed. More work is clearly
needed on this subject.

\subsection{The nucleon $\sigma_N$ term}

Experimental verifications of
Eq. \ref{SigmaN} (without the second term on its r.h.s.) have so far
been unsuccessful without assuming certain strange quark content of a
nucleon, which is discouraged by the OZI rule \cite{OZI}. A recent
review of the nucleon $\Sigma_N$ problem can be found in Ref.
\cite{GLS}. The experimental data for the scalar form factor of the
nucleon is known from experimental observations only on the
Cheng-Dashen point at $t=2m_\pi^2$. In order to make a
connection between this piece of experimental information with one
for the static scalar density $\sigma(0)$ that can be obtained,
within some specific models, from baryonic spectra. The extrapolation
of the scalar density from the Cheng-Dashen point to the $t=0$ static
point can be done using essentially model independent dispersion
relation method \cite{GLS}. A value of $\sigma(0) = 45$ MeV is
obtained from the experimental $\pi N$ scattering data.

The computations of the value of $\sigma(0)$, which relies upon the
information contained in the baryonic spectra,  can be classified into
two categories: (1) self-consistent mean field approximation without
the pion loop corrections \cite{Hats} (2) chiral perturbation theory
\cite{xPT} with loop corrections \cite{Gasser}.

One of the problems of the evaluation of $\sigma(0)$ from the
baryonic spectra is related to the possible non-linearity
in the current strange quark mass which renders the extracted value of
$\sigma(0)$ from baryonic spectra
unreliable. In the computations of the first category, the
non-linearity is large in Nambu Jona-Lasinio type of models
with contact interaction\cite{Hats}; it is found to be
small in models \cite{Meis}
where extended gluon propagators are used. A value of order $49.3$
MeV is obtained for $\sigma(0)$ in Ref. \cite{Hats}. It is already
somewhat larger than $45$ MeV obtained form the $\pi N$ scattering
data. The  computation given by Ref. \cite{Gasser}, which belongs to
the second category, obtains corrections to the $\sigma(0)$ term due to
pion loop contributions of order $10$ MeV. Ref. \cite{Sig-Delt}
computed the $\pi\Delta
$ components for a nucleon using bag and soliton models, which obtain
an additional $6$-$10$ MeV corrections.

Neither the computations given by Ref. \cite{Hats} nor the one given
by Ref. \cite{Gasser} is complete. The former ones miss the low lying
collective modes and the later ones contain large corrections to
$\sigma$, which imply self-consistent computations are needed.
Had we, for the purpose of estimation, simply add all the above
mentioned corrections up, we would have gotten a number for $\sigma(0)$
of order $68$ MeV. It is much larger than the one obtained from
$\pi N$ scattering data. It is even larger than the un-extrapolated
value of $\sigma(2m_\pi^2)$ of order $60$ MeV.

The problem is still not well understood.

It should be pointed out here that an additional source of correction
need to be considered to relate $\Sigma$, which is proportional to
the scalar density of the nucleon on the Cheng-Dashen point \cite{GLS},
to $\sigma(0)$ related to the static scalar density of a nucleon. In the
Hartree-Fock approximation, the static scalar density measured by
$\sigma$ is free of radiative corrections due to the fact that it
satisfies a ``gap equation'', which self-consistently adjust the
radiative corrections to $\sigma$ to zero (cancel) by changing its
value. This can be proven in the case that $\sigma$ is space-time
independent \cite{YING2}. We expect it to be true also for $\sigma(0)$.
This statement is not true if the matrix elements of the scalar
density operator ${1\over 2}\bar\Psi\Psi$ between states of different
4-momentum are taken. Various corrections due to the interaction have
to be considered.

$\Sigma = \sigma(2 m_\pi^2)$ term is written as $\Sigma =
\sigma_N(0) + \Delta_N$, $\sigma_N(0)=25 MeV$ can be obtained from the
baryon spectra \cite{GLS}. $\Delta_N$ contains various corrections due
to pionic modes \cite{GLS} to
the extrapolation. If there is diquark condensation in a nucleon,
there would be contributions to $\Delta_N$ that are proportional to
$\phi^2$ coming from the interaction terms. The existence of such a
term can be demonstrated at the quark level by considering a second
order correction to $\sigma_N$ due to the interaction, which has the
generic form $\delta \sigma_N^{(2)}\sim \mathopen{\langle p'\,|} T\left
(\Psi\bar
\Psi\right )\left (\bar\Psi \Gamma\Psi \right )\left (\bar\Psi \bar
\Gamma \Psi\right )\mathclose{|\, p\rangle}$, with $\Gamma$ and $\bar \Gamma$ a
pair of
the interaction vertices. This term contains $\phi^2$ contributions if
there is diquark condensation inside a nucleon \cite{diquark2}.
Detailed evaluation of these terms can not be proceeded
before a specific model is given. Here, we simply parameterize the
effects  of $\phi^2$ as
$\Sigma_N=\sigma_N(0)+\Delta_N^{(0)}+\beta_N\phi_N^2$, where
$\Delta_N^{(0)}$ is related to the total correction that has already
been considered in the literature (see, e.g., Ref. \cite{GLS}). The new
term depending on $\phi^2_N$ is due to the possible diquark
condensation inside a nucleon with an average strength measured by
$\sqrt{\phi_N^2}$. It is unclear at present whether the additional
term increases ($\beta_N<0$), decreases ($\beta_N>0$) or even
eliminates the above mentioned discrepancy.

\subsection{Deep inelastic scattering}

   The deep inelastic scattering (DIS) of charged leptons and neutrinos with
a nucleon and a nucleus provide another way of studying the constituent
quark (elementary excitation) inside that system. Of those relevant,
we emphasize in particular the change in behavior of the constituent
quark in a system with diquark condensation. It was discussed in Ref.
\cite{YING1} that diquark condensation spontaneously breaks the $U(1)$
symmetry corresponding to the baryon number conservation leading to a
superconducting phase inside the system. In this phase, the isoscalar
charge of the constituent quark, albeit does not disappear, is
spreaded, which can only be partially observed in the DIS. This phenomenon
is known in the study of superconducting condensed matter system
\cite{Nambu}.  The advantage of considering this particular aspect of
the constituent quark is related to the fact that there exist sum
rules for the quark based on the localization of the isoscalar charge,
which is violated in the superconducting phase.
In such a case, the electric charge of some \cite{YING2} of the
partons, which are the up and down quarks, have only isovector charge
$\pm 1/2$ in the limit of
large strength of diquark condensation. Therefore if there is a diquark
condensation that generates a localized
superconducting phase in a nucleon, the $F_1(x)$ and $F_2(x)$
structure functions should be written as
\begin{eqnarray}
   F_1(x) &=& {1\over 2} \sum_i \tilde Q^2_i f_i(x), \label{F_1equation}\\
   F_2(x) & = & \sum_i \tilde Q^2_i x f_i(x), \label{F_2equation}
\end{eqnarray}
where $i$ enumerates valence and sea quark components of the nucleon
and the effective charge $\tilde Q_i$, which should be replaced by
$\tilde C_V$ (for vector current operator) and $\tilde C_A$ (for axial
vector operator) in the neutrino DIS on a nucleon/nucleus,
is given in Table \ref{Table2}. The parameter $0 \le \alpha_i \le 1$ in
Table \ref{Table2} characterizes the strength of the superconducting phase.

In case of strong superconductivity, which means some of the
$\alpha_i$ is $0$, $F_2^N(x) \sim F_2^P(x)$. The experimental measurement of
$F_2^P(x)$ and $F_2^N(x)$ in the deep inelastic scattering experiments
shows such a tendency manifests in the violation of Gottfried sum rule
\cite{GOTT}. Perturbatively, the effects due to various flavor
violation is too small to account for the magnitude of violation.
One of the plausible interpretation is that there is
a flavor violation in the sea \cite{ISOv} due to mesonic
(pion and heavier mesons)
excitations.  The latest rather complete study (e.g. Szezurek and Speth
in Ref. \cite{ISOv}) still can not explain the violation. If we use the normal
electrical charges for the up and down quarks,
then there is a reduction of $f_i(x)$, which is interpreted as a violation of
the Gottfried sum rule.

In a large nucleus, the effects of diquark condensation are expected to be
enhanced due to the increase of the baryon density and the size of the
system; this in general produces the depletion of quark number
\cite{BFS} observed in the EMC effects in a nucleus, which is related
to a  (enhanced) violation of Gottfried
sum rule in the picture given here,  since the antiquark
components in a nucleus is known to be small in experimental observations
\cite{Eisele,Anti-quark}. Fermi motion and binding effects are not
enough to explain the EMC effects if proper normalization of baryon
number is carried out \cite{LLB} without rescaling.
It is generally accepted that the EMC effects are
still not well understood \cite{BPT}.

One of the decisive experiments, which can be used to test the picture
proposed for the violation of Gottfried sum rule and the EMC effects in this
paper, is the use of , instead of the charged lepton and nucleon/nucleus DIS,
the neutrino and nucleon/nucleus DIS. This is because the neutrino
couples only to the hadronic neutral current, which has smaller
isoscaler component of the vector current (see Table \ref{Table2}).
If the picture given here is true, then the magnitude of the violation
of the Gottfried sum rule and the EMC effects should be reduced in the
neutrino DIS on nucleon/nucleus.

The effective charges for the $\bar l + l\to hadrons$ production
processes are also given in Table \ref{Table2}. They are not affected
even if the hadrons produced in the final states
contains diquark condensation. Since the
quark-antiquark pair, which hadronizes into
hadrons in the final state, are current quarks created in the vacuum within
short
time period in the collision and
no diquark condensate is believed to be exist in the vacuum at the
present day conditions because it carries color.

\section{Summary}
\label{sec:Sum}

    A natural way of extending the PCAC relation beyond the conventional
one is presented in this study. For a relativistic fermionic system,
the extension presented here is unique if assumptions I, II are valid.
They are partially supported by the phenomenology.
The fact that there can only be two kinds of spontaneous
chiral symmetry breaking phases in nature that preserve flavor symmetry
considerably reduces the number of possibilities that one should consider.

A detailed study of the effects of these chiral
soft longitudinal modes and that of
spreading of the isoscalar part of the charge of constituent quarks
depends upon a more specific model for a nucleon. It
is beyond the scope of this paper, despite the fact that this study put strong
constraints on any such an attemp. It is an interesting topic to be
further explored.

\section*{Acknowledgment}

  The author would like to thank the Institute for Nuclear Theory at
the University of Washington for its hospitality during the completion of
the work. He would also like to thank Profs. E. M. Henley, R. K. Su and
G. J. Ni for discussions during the late stage of the work.



\begin{figure}
\caption{The  $\phi^2$ dependence of $a_\delta$.  $\Lambda$ is the
chiral symmetry breaking scale which is chosen to be the value of the
covariant cut off in the quark loop integral.
In order to obtain a smooth curve, the sharp cut off  in the Euclidean
momentum space integration is replaced by a smooth one. The shape
of the smooth cut off $F(p/\phi)$ is plotted on
the same graph with an arbitrarily chosen $\phi$.}
\label{Fig1}
\end{figure}

\begin{figure}
\caption{The  $\phi^2$ dependence of the quark-diquark coupling
constant $g_{\delta q}$. The
same smooth Euclidean momentum space cut off as the one used in Fig.
1 is used.
}
\label{Fig2}
\end{figure}

\begin{figure}
\caption{The  $\phi^2$ dependence of the Goldstone diquark decay
constant $f_\delta$. The same smooth Euclidean momentum space cut off
as the one used in Fig. 1 is used.}
\label{Fig3}
\end{figure}

\begin{figure}
\caption{The $\rho\pi$ reduced density of states in the pionic channel
          as a function of $q^2$ obtained from the fitting procedure
          for two values of the $\rho\pi\pi$ coupling constant.
          For comparison, the one loop ($\lambda=0$)
          reduced density of state is
          drawn with a dashed line.
          }
\label{Fig4}
\end{figure}



\begin{table}[h]
\caption{The results of fitting, where $g_A=1.26$, $g_{\pi NN}= 13.4$,
          $f_\pi=93.2$ MeV,
         $q_{min}^2=-0.6$ $\mbox{GeV}^2$ and $q_{max}^2=0.2$ $\mbox{GeV}^2$.
         The unit for all but $\lambda$ and $g_{\rho\pi\pi}$ is GeV.
         $\lambda$ and $g_{\rho\pi\pi}$ are dimensionless.
         The quantities with a star on top is chosen by physical
         considerations.\label{Table1}}

\begin{tabular}{|c|ccccccc|}
$g_{\rho\pi\pi}/4\pi$ & $\Lambda_{\pi NN}$ & $\lambda$ & $\sqrt{I_B}$ &
             $\sqrt{s_{R}^*}$ & $M_A^*$ & $\sqrt{s_{th}^*}$ &\\
\hline
1.0 & 0.94 &1.12 & 1.12 & 1.30 & 1.00 & 0.91 & \\
1.5 & 0.93 & 0.41 & 1.26 & 1.30 & 1.00 & 0.91 & \\
2.0 & 0.91 & -0.0023 & 1.34 & 1.30 & 1.00 & 0.91 & \\
2.5 & 0.89 & -0.25 & 1.27 & 1.30 & 1.00 & 0.91 & \\
2.9 & 0.88 & -0.39 & 1.30 & 1.30 & 1.00 & 0.91 & \\
3.5 & 0.86 & -0.56 & 1.35 & 1.30 & 1.00 & 0.91 & \\
\end{tabular}

\end{table}

\begin{table}[h]
\caption{The effective charges for the electro-weak couplings in the
  DIS between leptons and a nucleon/nucleus and in lepton-antilepton
  production of hadrons according to the standard
  model. $\tilde C_V$ and $\tilde C_A$ are coefficients of the vector
  and axial vector current operators
  in the hadronic weak neutral current operator.
  Here $\theta_W$ is the Weinberg angle, ``$hs$'' denotes
  hadrons, ``$u$'' and ``$d$''
  denotes up and down quark respectively. The corresponding
  charge for an anti-quark is just opposite. The value for $\alpha$ in
  the table depends on the color and the momentum fraction $x$
  of the corresponding quark if the
  flavor symmetry is preserved.
\label{Table2}}

\begin{tabular}{|c|cccc|}
Quark&$\tilde Q$($l + h\to l+ h$) &$\tilde C_A$($\nu + h\to \nu + h$)
&$\tilde C_V$($\nu + h\to \nu + h$)&\phantom{$\displaystyle ab \over c$}
 \\
\hline
 &&&& \\
 \large{u} & $\displaystyle \alpha {1\over 6} + {1\over 2}$ & $\displaystyle
{1\over 2}$ & $\displaystyle{1\over 2} -
  ({\alpha\over 3} + 1) sin^2\theta_W $&\\
 &&&& \\
 \large{d} &  $\displaystyle\alpha {1\over 6} - {1\over 2}$ & $\displaystyle
-{1\over 2}$ & $\displaystyle-{1\over 2} -
   ({\alpha\over 3} - 1) sin^2\theta_W$ &\\
 &&&& \\
\hline
Quark&\mbox{$Q$($\bar l+ l\to\overline{hs}+ hs)$} &
\mbox{$C_A$($\bar\nu+\nu\to\overline{hs}+ hs)$}
&\mbox{$C_V$($\bar\nu+\nu\to\overline{hs}+
hs$)}& \phantom{$\displaystyle ab \over c$}\\
\hline
 &&&& \\
\large{u} & $\displaystyle {2\over 3}$ & $\displaystyle {1\over 2}$ &
$\displaystyle
{1\over 2} - {4\over 3}sin^2\theta_W$ &\\
 &&&& \\
\large{d} & $\displaystyle-{1\over 3}$ & $\displaystyle-{1\over 2}$ &
$\displaystyle
-{1\over 2} +
{2\over 3}sin^2\theta_W $&\\
 &&&& \\
\end{tabular}

\end{table}


\begin{references}
\bibitem{diquark} The condensation of diquark can be defined by a localized
                  stabilizing non-vanishing $\phi_{c\mu}$ and
                  $\bar\phi^{\mu}_c$ auxiliary fields, which are introduced in
                  Ref. \cite{YING1}, inside the nucleon. By definition, all
                  quantum fluctuation in the quantum fields are included.
\bibitem{drive} A pseudo-scalar field can excite the axial vector Goldstone
                diquark inside a nucleon due to the presence of  baryonic
                matter.
\bibitem{diquark2} There is a
  normal ordered operators $:\bar\Psi_\alpha\bar\Psi_\beta\Psi_\gamma
  \Psi_\sigma:$ (with respect to the Nambu vacuum) in
  the Wick expansion of the time ordered product of the fermion fields
  in $\delta \sigma_N^{(2)}$. Part of their contributions to
  $\delta\sigma_N^{(2)}$, which are proportional to $\phi^2$, are
  non-vanishing if there is a diquark condensation inside a
  nucleon.
\end{references}
\end{document}